# Quest-V: A Virtualized Multikernel for High-Confidence Systems

Ye Li
Boston University
liye@cs.bu.edu

Matthew Danish
Boston University
md@cs.bu.edu

Richard West
Boston University
richwest@cs.bu.edu

**Abstract**

This paper outlines the design of 'Quest-V', which is implemented as a collection of separate kernels operating together as a distributed system on a chip. Quest-V uses virtualization techniques to isolate kernels and prevent local faults from affecting remote kernels. This leads to a high-confidence multikernel approach, where failures of system subcomponents do not render the entire system inoperable. A virtual machine monitor for each kernel keeps track of shadow page table mappings that control immutable memory access capabilities. This ensures a level of security and fault tolerance in situations where a service in one kernel fails, or is corrupted by a malicious attack. Communication is supported between kernels using shared memory regions for message passing. Similarly, device driver data structures are shareable between kernels to avoid the need for complex I/O virtualization, or communication with a dedicated kernel responsible for I/O. In Quest-V, device interrupts are delivered directly to a kernel, rather than via a monitor that determines the destination. Apart from bootstrapping each kernel, handling faults and managing shadow page tables, the monitors are not needed. This differs from conventional virtual machine systems in which a central monitor, or hypervisor, is responsible for scheduling and management of host resources amongst a set of guest kernels. In this paper we show how Quest-V can implement novel fault isolation and recovery techniques that are not possible with conventional systems. We also show how the costs of using virtualization for isolation of system services does not add undue overheads to the overall system performance.

*Categories and Subject Descriptors*  D.4.7 [*Operating Systems*]: Organization and Design

*General Terms*  Design, Reliability

*Keywords*  Virtualization, Multicore, Fault Isolation and Recovery

## 1. Introduction

Multicore processors are now ubiquitous in today's microprocessor and microcontroller industry. It is common to see two to four cores per package in embedded and desktop platforms, with server-class processors such as the Sun Niagara 3 having 16 cores and up to 8 hardware threads per core. Similarly, Intel's Single-chip Cloud Computer (SCC) supports 48 cores, and other manufacturers are following suit with an increase in core count driven in part by trade-offs in power and computational demands. Many emerging multicore processors now have support for hardware virtualization (e.g., Intel VT and AMD-V CPUs). Virtualization has re-emerged in the last decade as a way to consolidate workloads on servers, thereby providing an effective means to increase resource utilization while still ensuring logical isolation between guest virtual machines.

Hardware advances with respect to multicore technology have not been met by software developments. In particular, multicore processors pose significant challenges to operating system design [6, 8, 39]. Not only is it difficult to design software systems that scale to large numbers of processing cores, there are numerous micro-architectural factors that affect software execution, leading to reduced efficiency and unpredictability. Shared on-chip caches [21, 40], memory bus bandwidth contention [44], hardware interrupts [43], instruction pipelines, hardware prefetchers, amongst other factors, all contribute to variability in task execution times.

Coupled with the challenges posed by multicore processors are the inherent complexities in modern operating systems. Such complex interactions between software components inevitably lead to program faults and potential compromises to system integrity. Various faults may occur due to memory violations (e.g., stack and buffer overflows, null pointer dereferences and jumps or stores to out of range addresses [15, 28]), CPU violations (e.g., starvation and deadlocks), and I/O violations (e.g., mismanagement of access rights to files and devices). Device drivers, in particular, are a known source of potential dangers to operating systems, as they are typically written by third party sources and usually execute with kernel privileges. To address this, various researchers have devised techniques to verify the correctness of drivers, or to sandbox them from the rest of the kernel [35, 36].

In this paper, we present a new system design that uses both virtualization capabilities and the redundancy offered by multiple processing cores, to develop a reliable system that is resilient to software faults. Our system, called 'Quest-V' is designed as a multikernel [6], or distributed system on a chip. It uses virtualization techniques to isolate kernels on different cores of a multicore processor. Shadow page tables [3] securely isolate separate kernel images in physical memory. These page tables map each kernels 'guest' physical memory to host (or machine) physical memory. Changes to protection bits within shadow page tables can only be performed by a monitor associated with the kernel on the corresponding core. This ensures that any illegal memory accesses (e.g., write attempts on read-only pages) within a kernel are caught by the corresponding monitor. Our system has similarities to Barrelfish, in that it is a multikernel, while also using virtualization similar to systems such as Xen [5]. We differ from traditional virtualized systems [9] by only trapping into a monitor when a fault occurs. In all other situations, execution proceeds within the kernels or user-spaces on each core. Interrupts and communication directly involve kernels, without monitor intervention. Thus, for the most part, only memory virtualization using shadow paging is used for fault isolation.

We show how Quest-V does not incur significant operational overheads compared to a non-virtualized version of our system, simply called Quest, designed for SMP platforms. We observe that communication, interrupt handling, thread scheduling and system call costs are on par with the costs of conventional SMP systems, with the advantage that Quest-V can tolerate system component failures without the need for system reboots.

We show how Quest-V can recover from component failure using a network device driver example, whereby we detect software failure and restart a comparable driver either in a locally-recovered kernel or in an alternate kernel on another core. This serves as an example of the 'self-healing' characteristics of Quest-V, with online fault recovery being useful in situations where high-confidence (or high availability) is important. This is typically the case with many real-time and embedded mission-critical systems found in healthcare, avionics, factory automation and automotive systems, for example.

In the following two sections, we describe the Quest-V architectural design, introducing the goals first, followed by an overview of the system architecture. This is followed in Section 4 by an experimental evaluation of the system. Here, we show the overheads of online device driver recovery for a network device, along with the costs of using hardware virtualization to isolate kernels and system components. Section 5 describes related work, while conclusions and future work are discussed in Section 6.

## 2. Design Goals

The design and ongoing development of Quest-V is centered around three main goals: safety, predictability and efficiency. As part of our investigations into system safety, we have studied various methods for hardware and software fault isolation, including the use of type-safe languages [7, 20, 24, 25] and hardware features such as paging, segmentation [11, 38], and virtual machine support [3].

Quest-V is intended for safety-critical application domains, requiring high confidence. The National Coordination Office for Networking and Information Technology Research and Development (NCO/NITRD) defines a high confidence system as follows [17]:

*"A high confidence system is one in which the consequences of its behavior are well understood and predictable. It must withstand internal and external threats and must deal with naturally occurring hazards as well as malicious attacks from a sophisticated and well-funded adversary. Systems that employ HCS technologies will be resistant to component failure and malicious manipulation and will respond to damage or perceived threat by adaptation or reconfiguration."*

High confidence systems are found in numerous application domains, ranging from healthcare, avionics and automotive systems, to factory automation and robotics. With the emergence of off-the-shelf and low-power processors now supporting multiple cores and hardware virtualization, it seems appropriate that these will become commonplace within this class of systems. In fact, the ARM Cortex A15 is expected to feature virtualization capabilities, on a processing core typically designed for embedded systems.

The aim of this paper is to show that Quest-V can leverage virtualization technologies of modern processors, without undue overhead. We primarily use hardware virtualization support for memory sandboxing, partitioning the system into a collection of *sandbox kernels* that operate collectively as a distributed system on a chip. This differs from the traditional view of a hypervisor, or virtual machine monitor, which manages a number of guests that are logically isolated and operate as though they were mapped to separate physical machines. Under normal operation, Quest-V should only incur the overhead of address translation using shadow page tables.[1] Given hardware support for caching address translations from guest-virtual to machine-physical address, it is not necessarily a high cost to pay for the added protection of a system. In other respects, Quest-V allows sandbox kernels to communicate via shared memory channels that allow the exchange of messages, and via which services in one sandbox can access those in another. Similarly, physical devices are shareable across sandbox kernels and do not require a monitor to manage them on behalf of guests.

Finally, although this is not the primary focus of this paper, Quest-V is designed around *time as a first-class resource*. We have been developing the system to support virtual CPUs (VCPUs), which we will briefly describe in Section 3.4. Fundamentally, the goal here is to ensure that time management of CPU resources are accurately accounted on behalf of all threads, either those associated with conventional tasks or interrupt handlers. Interrupt handlers are schedulable in Quest-V and their priority is carefully aligned with any thread that may have initiated them, such as via a prior I/O request. We have shown this leads to a system that ensures *temporal isolation* between tasks and interrupts, so that no such entity ever executes for longer than its budgeted time in any window of real-time.

## 3. Quest-V Architecture Overview

As stated earlier, Quest-V relies on virtualization support for safety and isolation of system software components. The current implementation runs in 32-bit mode and is designed specifically for Intel x86 processors with virtual machine extensions. Plans are underway to support AMD and future ARM processors.

Unlike most of the traditional hypervisor and virtual machine systems, Quest-V primarily uses virtualization to implement an extra logical ring of memory protection into which *sandbox* address spaces are mapped. We define a sandbox domain to be a collection of one or more CPUs and the host physical memory that includes the local monitor, its sandbox kernel and all local applications. There is no need for CPU virtualization as the instructions of each sandbox kernel and its applications execute directly on the hardware. All x86 instructions that are traditionally troublesome for "trap and emulate" methods [29] of virtualization are addressed by hardware features available on modern processors. Similarly, complex I/O virtualization is not necessary with Quest-V since interrupts are directed to sandbox kernels, where they are handled without monitor intervention. Thus, Quest-V is essentially a distributed collection of sandbox kernels, each isolated using hardware managed page tables and restricted to a set of chosen CPUs.

The overhead of frequent switches between a hypervisor and a virtual machine (VM) are avoided to the maximum extent under Quest-V in order to improve performance. This is because a monitor does not exist to support and switch between multiple guest VMs. Instead, each sandbox kernel operates directly on the host hardware without monitor intervention, except for fault recovery, establishment of inter-sandbox communication channels, and initiating shadow page tables.

A high level overview of the Quest-V architecture design is shown in Figure 1. The details of each system component will be explained in later sections. A single hypervisor is replaced by a separate monitor for each sandbox kernel. In effect, memory virtualization becomes an integral design feature of Quest-V, to separate and recover from faults in system components, while providing performance that is otherwise comparable to a conventional SMP system. In fact, given the separation of data structures for each sandbox kernel, and the largely "share nothing" paradigm adopted by Quest-V, it is arguably more scalable than traditional SMP systems, although investigation of this is outside the scope of this paper. Here, we are more interested in describing how Quest-V is designed to address system component isolation and perform online fault recovery without compromising the entire system and requiring a full reboot. For us, Quest-V's objective is to meet the demands of high-confidence, safety-critical application domains.

---

[1] e.g., Hardware managed extended page tables on Intel x86 processors, or nested page tables on AMD processors.

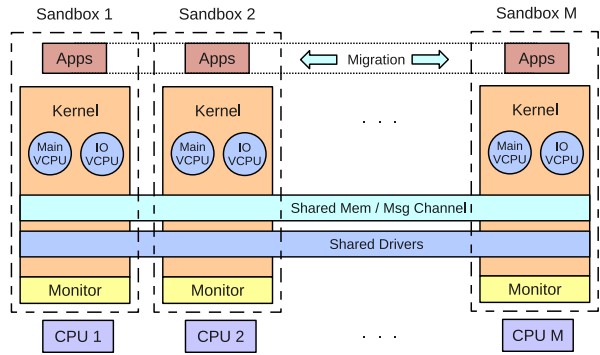

**Figure 1.** Quest-V Architecture Overview

The Quest-V architecture supports sandbox kernels that have both replicated and non-replicated services. That is, some sandboxes may have identical kernel functionality, while others partition various system components to form an asymmetric configuration. The extent to which functionality is separated across kernels is somewhat configurable in the Quest-V design. In our initial implementation, each sandbox kernel replicates most functionality, offering a private version of the corresponding services to its local applications. Certain functionality is, however, shared across system components. In particular, we share certain driver data structures across sandboxes [2], to allow I/O requests and responses to be handled locally. Using shared driver data structures, Quest-V allows *any* sandbox to be configured for corresponding device interrupts, rather than have a dedicated sandbox be responsible for all communication with that device. This greatly reduces the communication and control paths necessary for I/O requests from applications in Quest-V.

The multikernel model of Quest-V allows for software component failures in one kernel to be recovered either locally or remotely. In local recovery, a trap to the local monitor re-instates the faulty component in the sandbox kernel. Alternatively, in remote recovery, the faulty component can be replaced by a replicated or alternative component in a different sandbox. After recovery, a corresponding monitor relaunches the appropriate sandbox kernel to resume or restart the faulting service. The flexibility of service configuration and fault tolerance make Quest-V capable of providing reliable services in critical applications with a relatively low overhead.

It should be noted that, as part of the flexibility of the Quest-V design, each sandbox kernel can be configured to operate on a chosen subset of CPUs, or *cores*. In our current approach, we assume each sandbox kernel is associated with one physical core since that simplifies local (sandbox) scheduling and allows for relatively easy enforcement of service guarantees using a variant of rate-monotonic scheduling [23]. Notwithstanding, application threads can be migrated between sandboxes as part of a load balancing strategy.

### 3.1 Memory Layout

In the current implementation of Quest-V, different sandboxes have their own (host) physical memory region into which kernel images having almost identical functionality are mapped. Only the BIOS, certain driver data structures, and communication channels are shared across sandboxes, while all other functionality is privately mapped. A simplified layout of the memory partitioning scheme is illustrated in Figure 2.

---

[2] Only for those drivers that have been mapped as shareable between sandboxes.

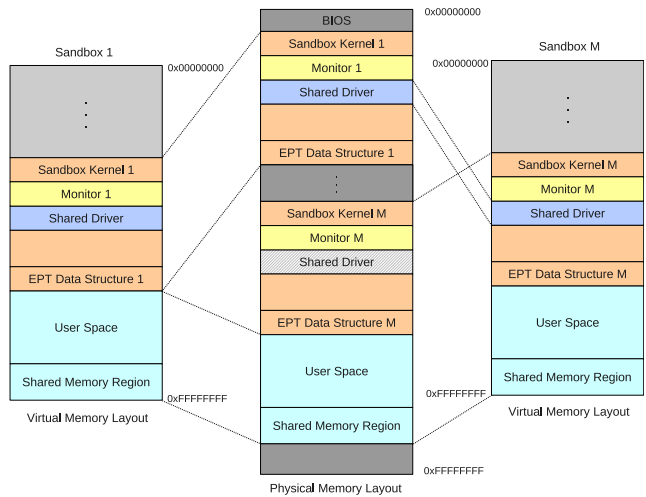

**Figure 2.** Quest-V Memory Layout

Each sandbox kernel image is mapped to physical memory after the region reserved for the system BIOS, beginning from the low 1MB. As stated earlier, a sandbox kernel can be configured to share one or more of its device drivers with any other sandboxes. Hardware resources can hence be partitioned for sandbox kernels with high flexibility and precision. Similarly, a user-level process which does not require strict memory protection can be loaded into a user space region accessible across sandboxes. This reduces the cost of process migration and inter-process communication. However, in the current Quest-V system, we do not support shared user-spaces for application processes, instead isolating them within the local sandbox. While this makes process migration more cumbersome, it prevents kernel faults in one sandbox from corrupting processes in others.

Regardless of shared or private user spaces, Quest-V supports inter-process communication (IPC) between processes in different sandboxes, using shared memory message passing channels. We reserve the highest addressable region of physical memory for this purpose. Finally, a memory region in each sandbox kernel is reserved for *extended page table* (EPT) data structures, and is used for memory protection, as explained in detail later in this section.

### 3.2 Hardware Virtualization Support

In order to enforce isolation between sandbox kernels, Quest-V utilizes the hardware virtualization support available in most of the current x86 and the next generation ARM processors to encapsulate each sandbox into a virtual machine. As with conventional hypervisors, Quest-V treats a guest VM domain as an extra ring of memory protection in addition to the traditional kernel and user privilege levels. However, instead of having one hypervisor for the whole system, Quest-V has one monitor running in the host domain for each sandbox as shown earlier in Figure 1. To avoid costly VM exits and entries, and associated overheads of monitor intervention, Quest-V grants maximum privilege to each VM domain. This allows each sandbox kernel to operate as though it were in a non-virtualized domain, except for memory address translation.

## 3.3 Hardware Assisted Shadow Paging

The isolation provided by memory virtualization requires additional steps to translate guest virtual addresses to host physical addresses. Modern processors with hardware support avoid the need for software managed shadow page tables, and they also support TLBs to cache various intermediate translation stages.

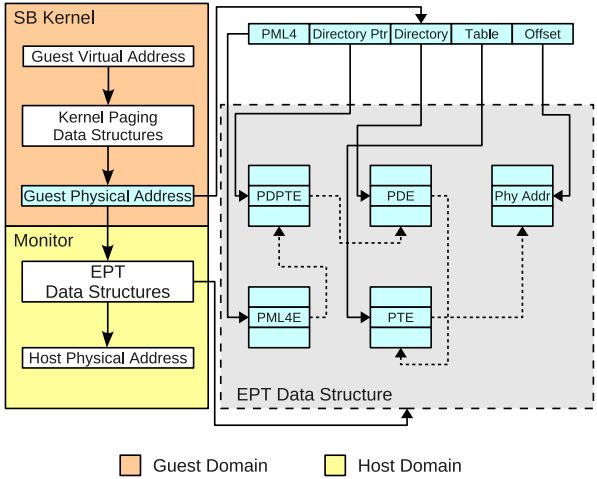

**Figure 3.** Extended Page Table Mapping

Figure 3 shows how address translation works for Quest-V guests (i.e., sandboxes) using Intel's extended page tables (EPTs). Specifically, each sandbox kernel uses its own internal paging structures to translate guest virtual addresses to guest physical addresses (GPAs). EPT structures are then walked by the hardware to complete the translation to host physical addresses (HPAs). The EPT structures are accessed by various bits of the guest physical address, starting with a pointer to a four-level outer table (the PML4 table). Each entry (PML4E) holds the host physical address of a page directory pointer table (PDPT), whose entries (PDPTEs) address a corresponding page directory table. The remaining two-levels involve directory and table entries to identify a host frame of memory, and the offset within that frame.

On modern Intel x86 processors with EPT support, address mappings can be manipulated at 4KB page granularity. This gives us a fine grained approach to isolate sandbox kernels and enforce memory protection. For each 4KB page we have the ability to set read, write and even execute permissions. Consequently, attempts by one sandbox to access illegitimate memory regions of another will incur an EPT violation, causing a trap to the local monitor. The EPT data structures are, themselves, restricted to access by the monitors, thereby preventing tampering by sandbox kernels.

By isolating sandboxes with EPT support, malfunctioning system components such as a faulty device driver will never be able to affect host physical memory regions that are either not present or marked as read-only in the local sandbox's EPT structures. If an EPT violation is detected, the monitor for the sandbox that triggered such an event can initiate fault recovery.

## 3.4 VCPU Scheduling Overview

In Quest-V, *virtual CPUs* (VCPUs) form the fundamental abstraction for scheduling and temporal isolation of the system. The concept of a VCPU is similar to that in virtual machines [3, 5], where a hypervisor provides the illusion of multiple *physical CPUs* (PCPUs) [3] represented as VCPUs to each of the guest virtual machines. VCPUs exist as kernel abstractions to simplify the management of resource budgets for potentially many software threads. We use a hierarchical approach in which VCPUs are scheduled on PCPUs and threads are scheduled on VCPUs.

A VCPU acts as a resource container [4] for scheduling and accounting decisions on behalf of software threads. It serves no other purpose to virtualize the underlying physical CPUs, since our sandbox kernels and their applications execute directly on the hardware. In particular, a VCPU does not need to act as a container for cached instruction blocks that have been generated to emulate the effects of guest code, as in some trap-and-emulate virtualized systems.

In common with bandwidth preserving servers [2, 13, 33], each VCPU, $V$, has a maximum compute time budget, $C_{max}$, available in a time period, $V_T$. $V$ is constrained to use no more than the fraction $V_U = \frac{C_{max}}{V_T}$ of a physical processor (PCPU) in any window of real-time, $V_T$, while running at its normal (foreground) priority. To avoid situations where PCPUs are otherwise idle when there are threads awaiting service, a VCPU that has expired its budget may operate at a lower (background) priority. All background priorities are set distinctly below those of foreground priorities to ensure VCPUs with expired budgets do not adversely affect those with available budgets.

Quest-V defines two classes of VCPUs: (1) *Main VCPUs* are used to schedule and track the PCPU usage of conventional software threads, while (2) *I/O VCPUs* are used to account for, and schedule the execution of, interrupt handlers for I/O devices. We now describe these two classes in more detail below.

***Main VCPUs.*** In Quest-V, Main VCPUs are by default configured as Sporadic Servers [32]. We use the algorithm proposed by Stanovich et al [34] that corrects for early replenishment and budget amplification in the POSIX specification. Fixed priorities are used rather than dynamic priorities (e.g., associated with deadlines) so that we can treat the entire system as a collection of equivalent periodic tasks scheduled by a rate-monotonic scheduler (RMS) [23]. Rate-monotonic analysis can then be used to ensure the utilization bound on any single PCPU does not exceed that required for a feasible schedule. In this approach, priorities are set inversely proportional to VCPU periods.

While a scheduling class defines a collection of threads and VCPUs, it is possible to assign different priorities to VCPUs (and also threads) within the same class. Moreover, multiple threads within the same class may share one or more VCPUs for their execution. By defaulting to a fixed priority scheme for scheduling VCPUs, we avoid the overhead associated with updating priorities dynamically, as would be the case if VCPUs had associated deadlines. While the least upper-bound on utilization for feasible schedules in static priority systems is often less than for dynamic priority systems, we consider this to be of lower importance when there are multiple PCPUs. With the emergence of multi- and many-core processors, it is arguably less important to guarantee the full utilization of every core than it is to provide temporal isolation between threads. In our case, the motivation is to provide temporal isolation between VCPUs supporting one or more threads.

Aside from temporal isolation of VCPUs, one additional factor in the design of Quest-V is the placement of VCPUs on PCPUs, to reduce microarchitectural resource contention. Guaranteeing a VCPU receives its bandwidth in a specified window of real-time does not guarantee that a thread using that VCPU will make efficient use of the corresponding CPU cycles. For example, a thread may stall on cache misses or memory bus bandwidth contention with other threads co-running on other cores. For this reason, Quest-V is being developed with a *performance monitoring* subsystem that inspects hardware performance counters to improve VCPU scheduling [40].

---

[3] We define a PCPU to be either a conventional CPU, a processing core, or a hardware thread in a simultaneous multi-threaded (SMT) system.

*I/O VCPUs.* For I/O VCPUs, we have considered several approaches for bandwidth preservation and scheduling. One approach is to use Sporadic Servers, but it is not clear what the most appropriate period should be to satisfy all I/O requests and responses. This is especially problematic when an I/O VCPU is shared amongst multiple tasks that issue I/O requests at different rates. While it is possible to dedicate a separate I/O VCPU to each task issuing I/O requests, so that individual bandwidth and rate requirements can be established, this adds overhead. Instead, it is preferable to use a single I/O VCPU for a given device that is shared across multiple tasks and, hence, Main VCPUs. Consequently, we allow shared I/O VCPUs to have fixed bandwidths but dynamic priorities that are inherited from the Main VCPUs of tasks responsible for I/O requests. Further details are described in a separate paper [12].

### 3.5 Inter-Sandbox Communication

Inter-sandbox communication in Quest-V relies on message passing primitives built on shared memory, and asynchronous event notification mechanisms using Inter-processor Interrupts (IPIs).

As mentioned in Section 3.1, there is a shared memory region for all the sandboxes in Quest-V in high physical memory. Shared memory is allocated using a separate physical memory manager and by default can be accessed by all the sandboxes. Some global resources that need to be shared among all sandbox kernels will have to be placed in this area. In the current Quest-V implementation, for instance, the global locks used to protect shared driver access reside in this region. Without isolation, shared memory should be allocated only when absolutely necessary. If *private* memory channels are needed between specific sandboxes, the local monitor is invoked to set up shadow page table mappings for the required access control.

By carefully controlling the degree of memory sharing, we can easily set up message passing channels between specific sandboxes and protect them from being compromised by any other sandboxes. A data structure resembling the concept of a *mailbox* will be set up on top of shared memory by each end of the communication channels for the purposes of exchanging messages. Asynchronous message passing in Quest-V can be supported by either polling for message reception, or using IPIs. We currently only support polled message exchanges using a status bit in each relevant mailbox to determine message arrival.

In addition to the basic message passing mechanism, Quest-V can also assign VCPUs to message passing threads, to constrain sending and receiving behavior in terms of CPU utilization and, hence, message throughput. This feature makes it possible for Quest-V to assign *priorities* and control message transmission rates for different communication channels, by binding sending and receiving threads to appropriately configured VCPUs. A benefit of this approach is that VCPUs scheduled using policies such as rate-monotonic scheduling can be guaranteed real-time shares of CPU cycles. This enables predictable communication between sandbox kernels, without unbounded delays for sending and receiving threads, while at the same time isolating their CPU usage from other time critical threads in the system.

### 3.6 Interrupt Distribution and I/O Management

In Quest-V, interrupts are delivered directly to sandbox kernels. Moreover, all sandbox kernels that share the use of a physical device all receive interrupts generated by that device. This avoids the need for interrupt handling to be performed in the context of a monitor as is typically done with conventional virtual machine approaches. Quest-V does not need to do this since complex I/O virtualization is not required. Instead, early demultiplexing in the sandboxed device drivers determines if subsequent interrupt handling should be processed locally. If that is not the case, the local sandbox simply discontinues further handling of the interrupt. We believe this to be less expensive than the cost of going through a dedicated coordinator as is done in Xen [5] and Helios [26], amongst others.

In order for the sandbox kernels to share a piece of hardware, we first need to find a way to deliver interrupts to them. This can be done either through interrupt broadcasting or IPIs. Most hardware platforms, including x86, provide an external programmable interrupt controller through which interrupts can be directed to specific processor cores and, hence, sandbox kernels. A logical structure of the Local and I/O APICs in Intel Xeon processors is shown in Figure 4.

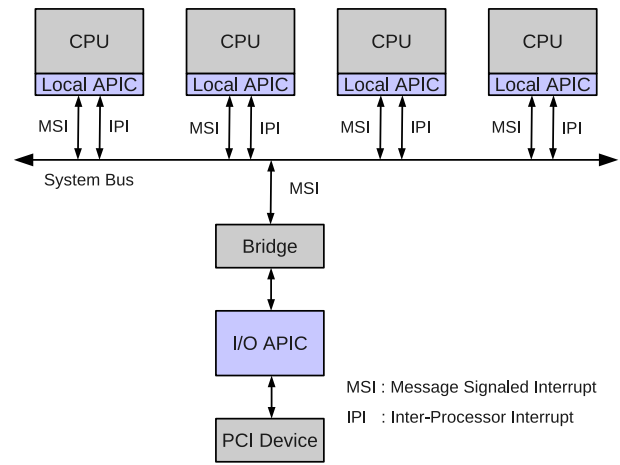

**Figure 4.** Local and I/O APICs in Intel Xeon Processors

If such a mechanism is not available or somehow prohibitive in certain scenarios, we can also direct interrupts to one physical processor and then use IPIs to simulate the broadcast. However, in the current Quest-V implementation, the I/O APIC is programmed to broadcast specific hardware device interrupts to all the sandboxes sharing that device. We also use this technique to re-route interrupts in case of fault recovery in a remote sandbox kernel, which will be discussed later.

Aside from interrupt handling, device drivers need to be written to support inter-sandbox sharing. Certain data structures have to be duplicated for each sandbox kernel, while others are protected by synchronization primitives. This is, in fact, a form of software-based I/O virtualization, although not as complex as that required by many hypervisors, which must manage transfers between devices and guests. The duplication of certain driver data structures, and the need for synchronization on shared data may impact the performance of hardware devices multiplexed between sandboxes. However, with the emergence of I/O virtualization technology, some hardware support for device sharing is now available. Technologies such as Intel's SR-IOV [19] could be useful for Quest-V. The basic idea here is combining early demultiplexing and separated DMA buffers to create a virtual view of the same device amongst those sandbox kernels using the device.

## 3.7 Fault Recovery

Fault detection is, itself, a topic worthy of separate discussion. In this paper, we assume the existence of techniques to identify faults. In Quest-V, faults are easily detected if they generate EPT violations, thereby triggering control transfer to a corresponding monitor. More elaborate schemes for identifying faults will be covered in our future work. In this section, we explain the details of how fault recovery is performed without requiring a full system reboot.

The multikernel approach adopted by Quest-V allows for fault recovery either in the local sandbox, where the fault occurred, or in a remote sandbox that is presumably unaffected. Upon detection of a fault, a method for passing control to the local monitor is required. If the fault does not automatically trigger a VM-exit, it can be forced by a fault handler issuing an appropriate instruction. [4] An astute reader might assume that carefully crafted malicious attacks to compromise a system might try to rewrite fault detection code within a sandbox, thereby preventing a monitor from ever gaining control. First, this should not be possible if the fault detection code is presumed to exist in read-only memory, which should be the case for the sandbox kernel text segment. This segment cannot be made write accessible since any code executing with the sandbox will not have access to the EPT mappings controlling host memory access. However, it is still possible for malicious code to exist in writable regions of a sandbox, including parts of the data segment. To guard against compromised sandboxes that lose the capability to pass control to their monitor as part of fault recovery, certain procedures can be adopted. One such approach would be to periodically force traps to the monitor using a *preemption timeout* [1]. This way, the fault detection code could itself be within the monitor, thereby isolated from any possible tampering from a malicious attacker or faulty software component. Many of these techniques are still under development in Quest-V and will be considered in our future work.

Assuming that a fault detection event has either triggered a trap into a monitor, or the monitor itself is triggered via a preemption timeout and executes a fault detector, we now describe how the handling phase proceeds.

*Local Fault Recovery.* In the case of local recovery, the corresponding monitor is required to release the allocated memory for the faulting components. If insufficient information is available about the extent of system damage, the monitor may decide to re-initialize the entire local sandbox, as in the case of initial system launch. Any active communication channels with other sandboxes may be affected, but the remote sandboxes that are otherwise isolated will be able to proceed as normal.

As part of local recovery, the monitor may decide to replace the faulting component, or components, with alternative implementations of the same services. For example, an older version of a device driver that is perhaps not as efficient as a recent update, but is perhaps more rigorously tested, may be used in recovery. Such component replacements can lead to system robustness through functional or implementation *diversity* [42]. That is, a component suffering a fault or compromised attack may be immune to the same fault or compromising behavior if implemented in an alternative way. The alternative implementation could, perhaps, enforce more stringent checks on argument types and ranges of values that a more efficient but less safe implementation might avoid. Observe that alternative representations of software components could be resident in host physical memory, and activated via a monitor that adjusts EPT mappings for the sandboxed guest.

*Remote Fault Recovery.* Quest-V also supports the recovery of a faulty software component in an alternative sandbox. This may be more appropriate in situations where a replacement for the compromised service already exists, and which does not require a significant degree of re-initialization. While an alternative sandbox effectively resumes execution of a prior service request, possibly involving a user-level thread migration, the corrupted sandbox can be "healed" in the background. This is akin to a distributed system in which one of the nodes is taken off-line while it is being upgraded or repaired.

In Quest-V, remote fault recovery involves the local monitor identifying a target sandbox. There are many possible policies for choosing a target sandbox that will resume an affected service request. However, one simple approach is to pick any available sandbox in random order, or according to a round-robin policy. In more complex decision-making situations, a sandbox may be chosen according to its current load. Either way, the local monitor informs the target sandbox via an IPI. Control is then passed to a remote monitor, which performs the fault recovery. Although out of the scope of this paper, information needs to be exchanged between monitors about the actions necessary for fault recovery and what threads, if any, need to be migrated.

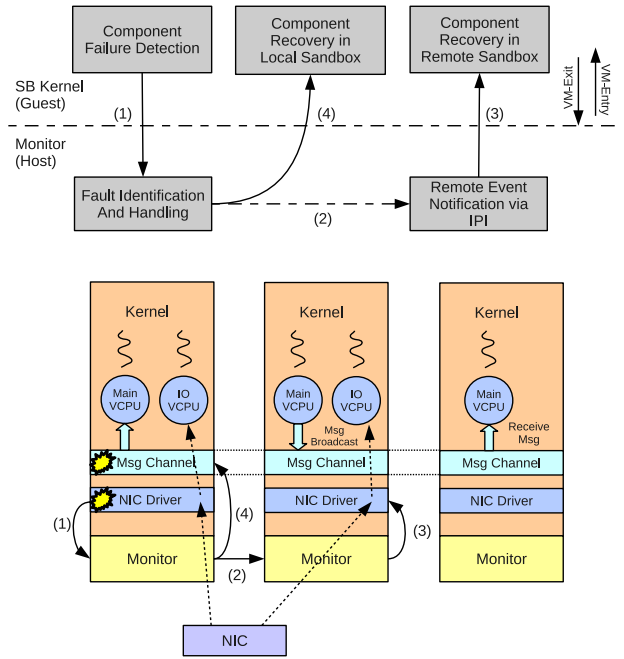

**Figure 5.** NIC Driver Remote Recovery

An example of remote recovery involving a network interface cased (NIC) driver is shown in Figure 5. Here, an IPI is issued from the faulting sandbox kernel to the remote sandbox kernel via their respective monitors, in order to kick-start the recovery procedures after the fault has been detected. For the purposes of our implementation, an arbitrary target sandbox was chosen. The necessary state information needed to restore service is retrieved from shared memory using message passing if available. In our simple tests, we assume that the NIC driver's state is not recovered, but instead the driver is completely re-initialized. This means that any prior in-flight requests using the NIC driver will be discarded.

The major phases of remote recovery are listed in both the flow chart and diagram of Figure 5. In this example, the faulting NIC driver overwrites the message channel in the local sandbox kernel. After receiving an IPI, the remote monitor resumes its sandbox kernel at a point that re-initializes the NIC driver. The newly selected sandbox responsible for recovery then redirects network interrupts to itself. Observe that in general this may not be necessary because interrupts from the network may already be broadcast and, hence, received by the target sandbox. Likewise, in this example, the target sandbox is capable of influencing interrupt redirection via an I/O APIC because of established capabilities granted by its monitor. It may be the case that a monitor does not allow such capabilities to be given to its sandbox kernel, in which case the corresponding monitor would be responsible for the interrupt redirection.

---

[4] For example, on the x86, the `cpuid` instruction forces a VM-exit.

When all the necessary kernel threads and user processes are restarted in the remote kernel, the network service will be brought up online. In our example, the local sandbox (with the help of its monitor) will identify the damaged message channel and try to restore it locally in step 4.

In the current implementation of Quest-V, we assume that all recovered services are re-initialized and any outstanding requests are either discarded or can be resumed without problems. In general, many software components may require a specific state of operation to be restored for correct system resumption. In such cases, we would need a scheme similar to those adopted in transactional systems, to periodically checkpoint recoverable state. Snapshots of such state can be captured by local monitors at periodic intervals, or other appropriate times, and stored in memory outside the scope of each sandbox kernel.

## 4. Experimental Evaluation

We conducted a series of experiments that compared Quest-V to both a Quest system without virtualization support, and also Linux. These experiments were intended to show the isolation, reliability and efficiency of the Quest-V design. The target machine was a Dell PowerEdge T410 server with an Intel Xeon E5506 processor, featuring 4 physical cores with both VMX and EPT support, along with 4GB RAM. An external PCIe Realtek R8169 network interface card was used for all network-related experiments.

### 4.1 Driver Fault Recovery

To demonstrate the system component fault recovery mechanism of Quest-V, we intentionally broke the Realtek R8169 NIC driver in one sandbox kernel, and tried to recover from the failure both locally in the current sandbox and remotely in another sandbox. The functioning status of the network sub-system was verified by a continuous stream of ICMP requests. By pinging the Quest-V test machine from another machine at a fixed rate during recovery, we measured the system down time and noted the extent to which ICMP packets were disrupted. This was possible by observing the received packet sequence numbers.

The total time spent on receiving 50 ICMP replies along with the down time caused by the driver fault recovery were recorded. To compare the efficiency of single component recovery versus a full system reboot, we also performed the same experiment for the case of a hardware reset (i.e., machine reboot). The results of the experiments are shown in Figure 6.

In both cases, the recovery begins after the reception of the twentieth ICMP echo reply. An ICMP echo request is sent every 500 milliseconds, as shown by the "ICMP Request" rate in the figure. As expected, the down-time of online recovery is much shorter than with a system reboot. The results for both local and remote recovery turn out to be very similar, so we only show the local recovery result in the figure.

As mentioned earlier, the network card driver local recovery process involves fault detection, followed by replacing and re-initializing the physical driver. Since fault detection is not in the scope of this paper, we triggered the fault recovery event manually by assuming an error occurred. Hence the down time involved only a driver replacement and re-initialization. However, since the NIC driver is closely coupled with the network interface, we also needed to bring down the old interface and start a new one for the sandbox.

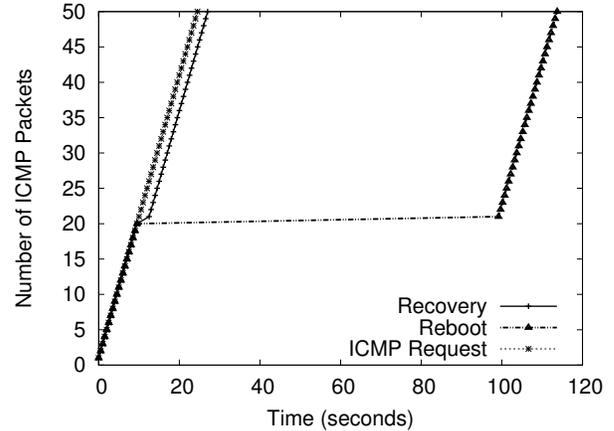

**Figure 6.** NIC Driver Recovery

| Phases | CPU Cycles | |
|---|---|---|
| | Local Recovery | Remote Recovery |
| VM-Exit | 707 | |
| Driver Switch | 12427 | N/A |
| IPI Round Trip | N/A | 1291 |
| VM-Enter | 823 | |
| Driver Re-initialization | 134244605 | |
| Network Re-initialization | 68750060 | |

**Table 1.** Overhead of Different Phases in Fault Recovery

The time Quest-V spent on different phases of the recovery is shown in Table 1. For most system components, we expect re-initialization to be the most significant recovery cost.

### 4.2 Forkwait Microbenchmark

To identify the costs of hardware virtualization in Quest-V, we generated a `forkwait` microbenchmark similar to that performed by Adams and Agesen [3]. In Quest-V, sandboxes spend most of their life-time in guest kernel mode, and system calls that trigger context switches will not induce VM-Exits to a monitor. Consequently, we tried to measure the overhead of hardware virtualization on normal system calls for Intel x86 processors. We chose the `forkwait` microbenchmark because it involves two relatively sophisticated system calls (`fork` and `waitpid`), involving both privilege level switches and memory operations.

Pseudocode for our `forkwait` program is outlined in Listing 1.

```
pid_t pid;
int i, status;
for (i = 0; i < 40000; i++) {
  pid = fork ();
  if (pid == 0) exit (0);
  else if (pid < 0) exit (1);
  else waitpid (pid, &status, 0);
}
```

**Listing 1.** Forkwait Pseudo Code

40000 new processes were forked in each set of experiments and the total CPU cycles were recorded. We then compared the performance of Quest-V against a version of Quest without hardware virtualization enabled, as well as a Linux 2.6.32 kernel in both 32- and 64-bit configurations. Results are shown in Figure 7.

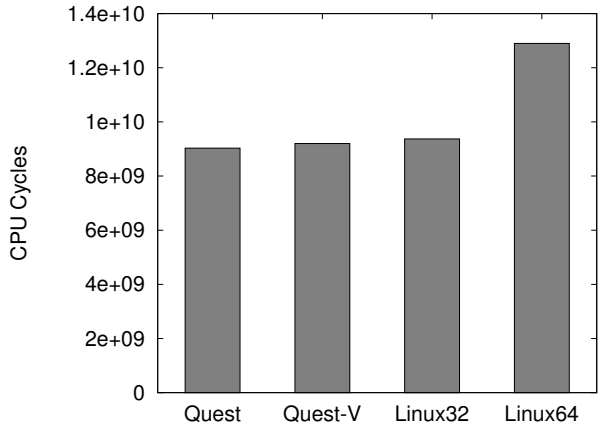

**Figure 7.** Forkwait Microbenchmark

We can see from the results that hardware virtualization does not add any obvious overhead to Quest-V system calls, and both Quest and Quest-V took less time than Linux to complete their executions. Even though this is not an entirely fair comparison because Quest is far less feature-rich than Linux, it should be clear that our approach of using hardware virtualization support for isolating software components is not prohibitive in the case of system calls.

### 4.3 Interrupt Distribution and Handling

Besides system calls, device interrupts also require control to be passed to a kernel. We therefore conducted a series of experiments to show the overheads of hardware virtualization on interrupt delivery and handling in Quest-V. For comparison, we recorded the number of interrupts that occurred and the total round trip time to process 30000 ping packets on both Quest and Quest-V machines. In this case, the ICMP requests were issued in 3 millisecond intervals from a remote machine. The results are shown in Figure 8.

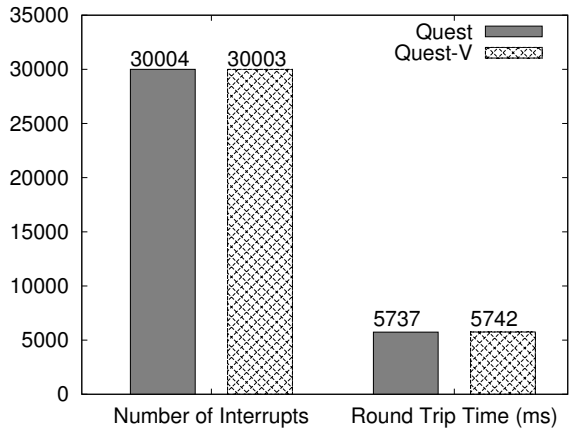

**Figure 8.** Interrupt Distribution and Handling Overhead

Notice that in Quest, all the network interrupts are directed to one core and in Quest-V, we broadcast network interrupts to all cores but only one core (i.e., one sandbox kernel) actually handles them. Each sandbox kernel in Quest-V performs early demultiplexing to identify the target for interrupt delivery, discontinuing the processing of interrupts that are not meant to be locally processed.

Consequently, the overhead with Quest-V also includes broadcasting of interrupts from the I/O APIC. However, we can see from the results that the performance difference between Quest and Quest-V is almost negligible, meaning neither hardware virtualization nor broadcasting of interrupts is prohibitive. It should be noted that Quest-V does not require intervention of a monitor to process interrupts. Instead, interrupts are directed to sandbox kernels according to rules setup in corresponding virtual machine control structures.

### 4.4 Inter-Sandbox Communication

The messaging passing mechanism in Quest-V is built on shared memory, and since we do not support NUMA for now, its performance is mostly affected by memory access, cache and scheduling costs. Instead of focusing on memory and cache optimization, we tried to study the impact of scheduling on message passing in Quest-V.

Specifically, we setup two kernel threads in two different sandbox kernels and assigned a VCPU to each of them. One kernel thread used a 4KB shared memory message passing channel to communicate with the other thread. In the first case, the two VCPUs were the highest priority with their respective sandbox kernels. In the second case, the two VCPUs were assigned lower utilizations and priorities, to identify the effects of VCPU parameters (and scheduling) on the message sending and receiving rates. In both cases, the time to transfer messages of various sizes across the communication channel was measured. Note that the VCPU scheduling framework ensures that all threads are guaranteed service as long as the total utilization of all VCPUs is bounded according to rate-monotonic theory [23]. Consequently, the impacts of message passing on overall system performance can be controlled and isolated from the execution of other threads in the system.

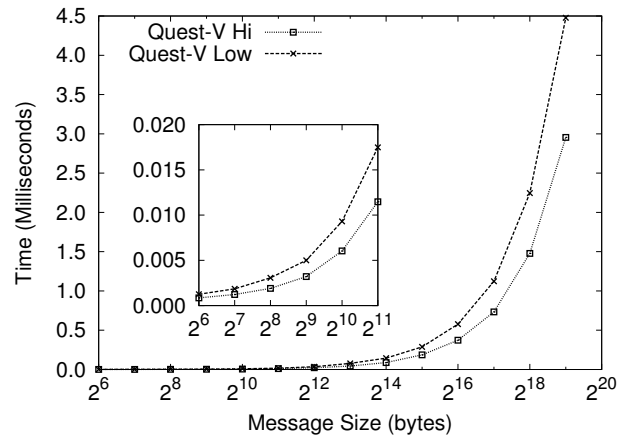

**Figure 9.** Message Passing Microbenchmark

Figure 9 shows the time spent exchanging messages of various sizes. Note that the x-axis in the figure is plotted using a log scale. *Quest-V Hi* is the plot for message exchanges involving high-priority VCPUs having utilizations of 50% for both the sender and receiver. *Quest-V Low* is the plot for message exchanges involving low-priority VCPUs having utilizations of 40% for both the sender and receiver. As can be seen, the VCPU parameters have an effect on message transfer times.

In our experiments, the time spent for each size of message was averaged over a minimum of 5000 trials to normalize the scheduling overhead. The communication costs grow linearly with increasing message size, because they include the time to write the message to, and read it from, memory. Discounting the memory read and write costs, the overhead is a constant to identify the shared memory address for a message.

### 4.5 Isolation

To demonstrate the fault isolation and recovery features of Quest-V, we designed a scenario that includes both message passing and network service across 4 different sandboxes. Specifically, sandbox 1 has a kernel thread that sends messages through private message passing channels to sandbox 0, 2 and 3. Each private channel is shared only between the sender and specific receiver, and is guarded by shadow page tables. In addition, sandbox 0 also has a network service running that handles ICMP echo requests. After all the services are up and running, we manually break the NIC driver in sandbox 0, overwrite sandbox 0's message passing channel shared with sandbox 1, and try to wipe out the kernel memory of other sandboxes to simulate a driver fault. After the driver fault, sandbox 0 will try to recover the NIC driver along with both network and message passing services running in it. During the recovery, the whole system activity is plotted in terms of message reception rate and ICMP echo reply rate in all available sandboxes and the results are shown in Figure 10.

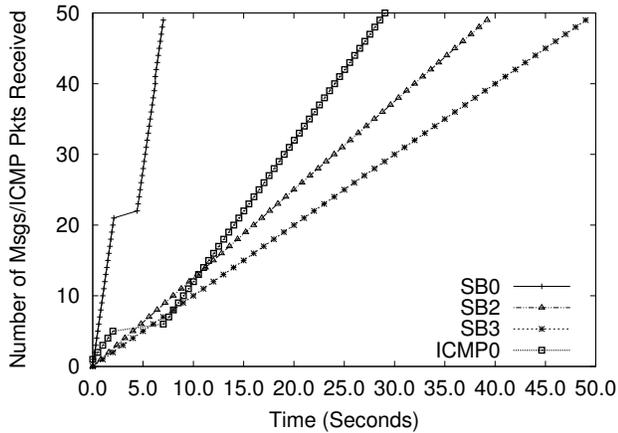

**Figure 10.** Sandbox Isolation

In the experiment, sandbox 1 broadcasts messages to others at 50 millisecond intervals, while sandbox 0, 2 and 3 receive at 100, 800 and 1000 millisecond intervals. Also, another machine in the local network sends ICMP echo requests at 500 millisecond intervals to sandbox 0.

We can see from the results that an interruption of service happened for both message passing and network packet processing in sandbox 0, but all the other sandboxes were unaffected. This is because each sandbox has its physical memory separated from the others through hardware assisted shadow paging (EPT in x86). When the "faulty" driver in sandbox 0 tries to overwrite memory of the other sandboxes, it simply traps into the local monitor because of a memory violation. Consequently, the only memory that the driver can wipe out is only the writable memory in sandbox 0. Hence all the monitors and all the other sandboxes will remain protected from this failure.

### 4.6 Shared Driver Performance

Quest-V supports device driver sharing, as mentioned in Section 3.6. A simple I/O virtualization strategy allows a physical driver to be mapped into all sandboxes requiring access to the same device, with each sandbox handling device interrupts locally. This differs from traditional I/O virtualization techniques in which interrupts are first handled in a monitor. An example shared driver in Quest-V might be for a single NIC device, providing a separate virtual interface for each sandbox requiring access. This allows for each sandbox to have its own IP address and even a virtual MAC address for the same physical NIC.

We compared the performance of our shared driver design to the I/O virtualization adopted by Xen 4.1.2 and VMware Player 4.0. We used an x86_64 root-domain (Dom0) for Xen, based on Linux 3.1. VMPlayer was hosted by a Fedora Core 16 64-bit Linux. We also used Ubuntu Linux 10.04 (32-bit kernel version 2.6.32) as a base case system without I/O virtualization. In all cases, a flood of 30000 ICMP echo requests were delivered to each guest, or system, on a target host with a Realtek NIC. The round-trip response time was recorded on each machine acting as the source of the ICMP ping packets. Figure 11 shows the results for the target host running different OS configurations in separate ping flood experiments.

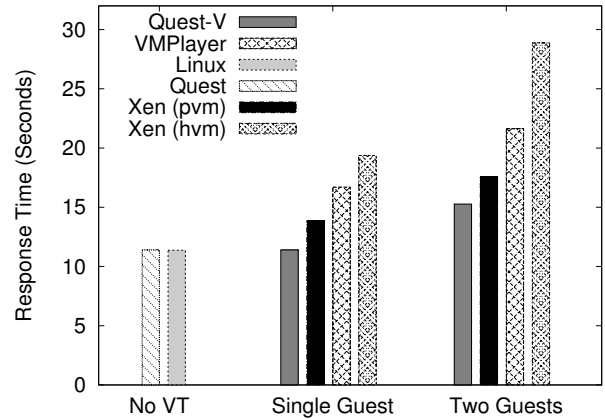

**Figure 11.** Shared NIC Driver Performance

The "No VT" case compares Quest without virtualization against Ubuntu Linux. Both systems show almost identical performance. In the "Single Guest" case, one sandbox kernel accesses the NIC in Quest-V. For comparison, we show results for a single Linux VM hosted by VMPlayer, and a single Linux user-domain (DomU) hosted by Xen. Two separate experiments were performed using Xen: "Xen (pvm)" featured guest paravirtualization, while "Xen (hvm)" used hardware virtualization. For both VMPlayer and Xen, we again used Ubuntu Linux 10.04 guests. Results indicate the virtualization overhead of Quest-V is less than for VMPlayer and Xen. This is the case even when Xen uses paravirtualization.

Finally, to show the effects of device sharing, we first have two sandboxes in Quest-V receiving ping packets via separate IP addresses. Here, two separate hosts act as the source of ping packets delivered to the target machine. For comparison, two Linux VMs hosted by VMPlayer, and two Linux DomUs in Xen also receive network ping traffic via the same physical NIC. The average round-trip response time for both guests (or sandboxes) is recorded in the "Two Guests" case. Results show that device sharing affects the performance of all systems but Quest-V has lower overhead than both VMPlayer and Xen. These initial results suggest Quest-V's shared driver approach is not prohibitive to performance.

### 4.7 Discussion

As stated earlier, Quest-V uses hardware virtualization primarily for memory isolation of sandbox kernels, for heightened dependability. The experiments show that the overheads of using virtualization in this manner are not significant compared to a conventional non-virtualized SMP system. In fact, Quest-V conducts most of its operation without intervention of a monitor, except to establish shadow page tables, launch sandbox kernels, and handle faults. Only the added address translation costs are incurred under normal operation.

For situations where a conventional system might be compromised and require a full reboot, Quest-V only has to perform online recovery of the sandbox kernel that is at fault. Experiments show that online fault recovery is possible with Quest-V and the approach is potentially applicable to a wider range of situations than we have discussed. For mission-critical applications, where the cost of reboot and loss of system functionality for this period of time is potentially disastrous, Quest-V is a possible solution. Online recovery allows unaffected parts of the system to remain functional, without any down-time.

Although we have only shown a relatively simple example for online recovery, using a ICMP network-level experiment that is largely stateless, it is still possible to support stateful recovery schemes. Such schemes would likely have to checkpoint state, in the same way transactional systems work [30], and roll-back to the last checkpoint in their recovery. To prevent the checkpointed state itself being corrupted by a fault, Quest-V could take advantage of memory regions accessible only to the local monitor of each sandbox. In this way, checkpoints would require VM-exit transitions from sandbox kernels to monitors to take snapshots of execution state. While VM-exits and entries incur almost a thousand clock cycles or more of overhead, this can be ameliorated by less frequent checkpoints, at the cost of potentially further rollback during system recovery.

## 5. Related Work

The concept of a multikernel is featured in Barrelfish[6], which has greatly influenced our work. Barrelfish replicates system state rather than sharing it, to avoid the costs of synchronization and management of shared data structures. The overall design is meant to be hardware-neutral with only the message transport mechanism and hardware interface being architecturally specific. As with Quest-V, communication between kernels is via explicit message passing, using shared memory channels to transfer cache-line-sized messages. In contrast to Barrelfish, Quest-V uses virtualization mechanisms to partition separate kernel services as part of our goal to develop high-confidence systems. Reliability and fault tolerance is key to the design of Quest-V, which we aim to use in real-time, safety-critical application domains. Shadow page table structures, such as those supported by hardware virtualization, ensure immutable protection capabilities between kernel components.

Systems such as Hive [10] and Factored OS (FOS) [39] also take the view of designing a system as a distributed collection of kernels on a single chip. FOS is primarily designed for scalability on manycore systems with potentially 100s to 1000s of cores. Each OS service is factored into a set of communicating servers that collectively operate together. In FOS, kernel services are partitioned across spatially-distinct servers executing on separate cores, avoiding contention on hardware resources such as caches and TLBs. Quest-V differs from FOS in its primary focus, since the former is aimed at fault recovery and dependable computing. Moreover, Quest-V manages resources across both space and time, providing real-time resource management that is not featured in the scalable collection of microkernels forming FOS.

Hive [10] is a standalone OS that targets features of the Stanford FLASH processor to assign groups of processing nodes to *cells*. Each cell represents a collection of kernels that communicate via message exchanges. The whole system is partitioned so that hardware and software faults are limited to the cells in which they occur. Such fault containment is similar to that provided by virtual machine sandboxing, which Quest-V relies upon. However, unlike Quest-V, Hive enforces isolation using special hardware *firewall* features on the FLASH architecture.

There have been several notable systems relying on virtualization techniques to enforce logical isolation and implement scalable resource management on multicore and multiprocessor platforms. Disco [9] is a virtual machine monitor (VMM) that was key to the revival in virtualization in the 1990s. It supports multiple guests on multiprocessor platforms. Memory overheads are reduced by transparently sharing data structures such as the filesystem buffer cache between virtual machines.

Xen[5] is a subsequent VMM that uses a special driver domain and (now optional) paravirtualization techniques [41] to support multiple guests. In contrast to VMMs such as Disco and Xen, Quest-V operates as a single system with sandbox kernels potentially implementing different services that are isolated using memory virtualization. Aside from a monitor on each core being used for establishing shadow pages and handling faults, the sandbox kernels of Quest-V are given direct access to devices and physical CPUs. This greatly reduces the cost of virtualization. For example, it allows I/O requests to be initiated and delivered directly to any core, as in a typical SMP system without I/O virtualization overheads.

Cellular Disco [16] extends the Disco VMM with support for hardware fault containment. As with Hive, the system is partitioned into cells, each containing a copy of the monitor code and all machine memory pages belonging to the cell's nodes. A failure of one cell only affects the VMs using resources in the cell. In effect, Cellular Disco treats a shared memory multiprocessor as a *virtual cluster*. Quest-V does not focus explicitly on hardware fault containment but its system partitioning into separate kernels means that it is possible to support such features. While we do not focus on this aspect, it is also possible for Quest-V to define domains, or cells, in which a monitor is established for a cluster of cores rather than just one. The design of Quest-V is not inherently tied to one core per sandbox kernel.

Another system that adopts the concept of multiple kernels is Helios [26]. Helios features *satellite kernels* that execute on heterogeneous platforms, including graphics processing units, network interface cards, or specific NUMA nodes. Applications and services can be off-loaded to special purpose devices to reduce the load on a given CPU. Helios builds upon Singularity [18] and all satellite microkernels communicate via message channels. Device interrupts are directed to a *coordinator* kernel, which restricts the location of drivers. In Quest-V, I/O APICs are allowed to broadcast interrupts to all accessible Local APICs, where early-demultiplexing [37] drivers detect whether the interrupt should be subsequently handled or discarded in the context of a time-budgeted I/O VCPU.

Helios, Singularity, and the *Sealed Process Architecture* [18] enforce dependability and safety using language support based on C#. Singularity, in particular, enforces isolation between software components mapped to *software isolated processes*, which are logically separated from an underlying microkernel. In Quest-V, virtualization techniques are used to isolate software components. While this may seem more expensive, we have seen on modern processors with hardware virtualization support that this is not the case.

In other work, Corey[8] is a library OS providing an interface similar to the Exokernel[14], and which attempts to address the bottlenecks of data sharing across modern multicore systems. Corey provides several abstractions, most notably *shares*, *address ranges* and *processing cores* that are managed by library OSes and which can be space-partitioned between applications. The kernel core feature of Corey provides the ability to allow applications to dedicate cores to kernel functions and data. A kernel core can manage hardware devices and execute system calls sent from other cores.

Multiple application cores then communicate with the kernel core via shared memory IPC. This is a bit different from Quest-V in that Quest-V asks multiple sandboxes to share a system component (shared drivers) if necessary. However, the flexibility of Quest-V allows it to be configured either way. We chose the former way to eliminate the overhead of IPC and control flow transfer that involves a monitor. Notwithstanding, with driver support it is possible to adopt the same approach as Corey.

Finally, Quest-V has similarities to systems that support self-healing, such as ASSURE [31] and Vigilant [27]. Such self-healing systems contrast with those that attempt to verify their functional correctness before deployment. seL4 [22] attempts to verify that faults will never occur at runtime, but as yet has not been developed for platforms supporting parallel execution of threads (e.g., multicore processors). Regardless, verification is only as good as the rules against which invariant properties are being judged, and as a last line of defense Quest-V is able to recover at runtime from unforeseen errors.

## 6. Conclusions and Future Work

This paper describes a system, called Quest-V, which uses virtualization to isolate system software components for improved reliability. We use hardware virtualization to isolate separate kernel images and their applications on the different cores of a multicore processor. This leads to a system design that resembles a distributed system on a chip. Shadow page tables control the access rights to host physical memory for each kernel. This ensures that faults in one kernel do not affect the memory spaces of others, thereby ensuring fault isolation. Our goal is to support high-confidence systems that must ensure survivability in the face of faults and potential attacks that may compromise system integrity. This is particularly important in real-time safety-critical domains.

With our approach, we are able to implement different services across sandbox kernels. Additionally, through a fault detection and recovery system, we are able to replace faulty services with alternative implementations in another sandbox kernel. We describe several experiments to show that the overheads of using virtualization technology for fault isolation is not prohibitive. Interrupts and communication between kernels do not require intervention of a monitor, as is typically the case with traditional virtual machine systems. Similarly, device drivers execute within sandbox kernels as they would in conventional non-virtualized systems. That is, they do not have to be specifically tailored to a virtual domain, providing a different interface to that of the underlying hardware, as is the case in systems using paravirtualization [5, 41].

A case study, using a network device driver shows how potential faults in one sandbox kernel can be recovered with the help of a local monitor. The costs of fault recovery are significantly less than would be the case with a full system reboot. We provide a simple example using network packet transfer, to demonstrate how a faulty driver can be restored in the original sandbox or even an alternative sandbox. Quest-V allows alternative versions of services to exist in different sandbox kernels, so that if one such service is compromised (e.g., via a malicious attack from a network intruder) a different version may be used as a replacement during recovery. This, then, allows for multiple versions of services to exist, perhaps trading performance for reliability.

This paper assumes the existence of a fault detection mechanism, which transfers control to a monitor. In future work, we will investigate techniques for fault detection and recovery that allow services to be resumed for stateful tasks. This would require recovery techniques similar to those in transactional systems [30].

Finally, although Quest-V is a working multikernel that uses hardware virtualization for protection and fault recovery, it is still fairly limited compared to more mature systems such as Linux. Quest-V lacks the rich APIs and libraries commonly found in modern systems, which restricts our ability to draw direct comparisons against current alternatives. Our future plans are to develop Quest-V with more extensive features, while continuing to leverage multicore and hardware virtualization capabilities. We will use hardware performance counters to improve the efficient and predictable management of hardware resources, including shared caches and NUMA interconnects. We hope Quest-V will pave the way for future highly reliable, self-healing systems. Along with complementary techniques, such as static verification, we will continue our development of a system that focuses on three goals: safety, predictability and efficiency.

## Acknowledgment

This work is supported in part by NSF Grant #1117025. The authors would like to thank Gary Wong for co-development of the original Quest system.